# DIELECTRIC MATRIX AND PLASMON DISPERSION IN STRONGLY COUPLED ELECTRONIC BILAYER LIQUIDS

Kenneth I. Golden* and Hania Mahassen
*Department of Mathematics and Statistics*
*University of Vermont, Burlington, Vermont 05405, USA*

Gabor J. Kalman
*Department of Physics*
*Boston College, Chestnut Hill, Massachusetts 02467, USA*

Gaetano Senatore
*Dipartimento di Fisica Teorica*
*Università di Trieste, Strada Costiera 11, 34014 Trieste, Italy*
*INFM-Democritos National Simulation Center, Trieste, Italy*

F. Rapisarda†
*Dipartimento di Fisica Teorica*
*Università di Trieste, Strada Costiera 11, 34014 Trieste, Italy*

# ABSTRACT

We develop a dielectric matrix and analyze plasmon dispersion in strongly coupled charged-particle bilayers in the $T = 0$ quantum domain. The formulation is based on the classical quasi-localized charge approximation (QLCA) and extends the QLCA formalism into the quantum domain. Its development, which parallels that of 2D companion paper [Phys. Rev. E **70**, 026406 (2004)] by three of the authors, generalizes the single-layer scalar formalism therein to a bilayer matrix formalism. Using pair correlation function data generated from diffusion Monte Carlo simulations, we calculate the dispersion of the in-phase and out-of-phase plasmon modes over a wide range of high-$r_s$ values and layer spacings. The out-of-phase spectrum exhibits an exchange-correlation induced long-wavelength energy gap in contrast to earlier predictions of acoustic dispersion softened by exchange-correlations. The energy gap is similar to what has been previously predicted for classical charged-particle bilayers and subsequently confirmed by recent molecular dynamics computer simulations.

## I. INTRODUCTION

Over the past two decades, interest in strongly coupled multilayer plasmas has been stimulated by a confluence of experimental and theoretical activities in the areas of strongly coupled plasma physics and condensed matter physics. Foremost in the area of strongly coupled plasma physics are the pioneering NIST/Boulder experiments where laser-cooled classical ions in a cryogenic trap spontaneously organize themselves into layered one-component plasma (OCP) structures in highly correlated liquid and solid phases [1] (see Refs. [2] and [3] for related molecular dynamics (MD) and theoretical studies). In the area of condensed matter plasmas, there has been considerable interest in the fabrication of high-$r_s$ – multiple quantum well structures of parallel charged-particle layers [4]. So far, advances in modern semiconductor nanotechnology have made it possible to routinely fabricate high-mobility single two-dimensional (2D) layers in a strongly correlated Coulomb liquid phase at temperatures well below and comparable with the Fermi temperature. Experimental studies pursued in this domain include (i) recent measurements of the spin susceptibility of a 2D electron system over a wide range of $r_s$ values [5], (ii) measurements of the compressibility of a 2D hole liquid as it crosses the metal-insulator transition boundary [6], and (iii) recent inelastic light-scattering-based measurements of plasmon dispersion in high-quality low-density 2D electron liquids [7]. One would expect that similar high-$r_s$ experimental techniques will become available to bilayers and double quantum wells as well in the near future.

This paper addresses the problem of longitudinal collective mode dispersion in the strongly coupled zero-temperature electronic bilayer liquid. The symmetric bilayer is modeled as two equal-density $(n_1 = n_2 = n = N/A)$ monolayers of mobile electrons (or holes), each layer immersed in its own two-dimensional (2D) uniform neutralizing background of opposite charge. The 2N charges occupy the large but bounded area $A$ in the planes $z = 0$ and $z = d$ of a Cartesian coordinate system, $d$ being the interlayer spacing. The interaction potentials for the symmetric charged-particle bilayer are

$$\phi_{11}(r) = \phi_{22}(r) = e^2/(\varepsilon_s r), \qquad \phi_{12}(r) = e^2/\left(\varepsilon_s \sqrt{r^2 + d^2}\right),$$

$$\phi_{11}(q) = \phi_{22}(q) = 2\pi e^2/(\varepsilon_s q), \qquad \phi_{12}(q) = \left[2\pi e^2/(\varepsilon_s q)\right]\exp(-qd), \quad (1)$$

$r$ and $q$ being the in-layer separation distance and wave number, respectively. The parameter $r_s = a/a_B$ is the customary measure of the in-layer coupling strength in the zero-temperature quantum domain, $a = 1/\sqrt{\pi n}$ being the 2D Wigner-Seitz radius and $a_B = \hbar^2 \varepsilon_s / me^2$ the effective Bohr radius; $\varepsilon_s$ is the dielectric constant of the substrate.

Historically, the longitudinal collective mode structure of layered electron gases was studied first in relation to the type-I superlattice in the hydrodynamic approximation [8] and in the random-phase approximation (RPA) [9, 10]. It was found that the RPA longitudinal mode structure consists of an isolated 3D bulk plasmon and a band of acoustic modes [8, 9]. The RPA mode structure of the charged-particle bilayer was studied by Das Sarma and Madhukar and by Santoro and Giuliani [11]. The longitudinal mode structure consists of an in-phase (+) mode (where the two layers oscillate in unison) and an out-of-phase $(-)$ mode (where the oscillations of the two layers exhibit a 180° phase difference). Within the RPA, the in-phase plasmon has the typical $\omega_+(q \to 0) \sim \sqrt{q}$ corresponding to the optical mode of an isolated 2D layer of density 2n, while the out-of-phase plasmon is acoustic, $\omega_-(q \to 0) \sim q$ [11]. To better facilitate comparison with experiments on the layered electron gas in the $r_s < 1$ (high-density) regime, Jain and Allen [10(a)] subsequently derived the RPA plasmon dispersion relations for semiconductor multilayer systems consisting of a *finite* number of equal-density layers. Their theoretical predictions [10] about the RPA dispersion of the out-of-phase acoustic plasmons are in good agreement with measurements from Raman scattering experiments in the low-$r_s$ regime [12].

In the strong coupling regime, the RPA is no longer applicable. In the high-temperature classical domain, two substantially different theoretical approaches have been proposed: The first is the conventional 3D Singwi-Tosi-Land-Sjolander (STLS) [13] approach adapted to the calculation of the dynamical dielectric matrix for the type-I superlattice [14]. The second approach invokes the quasilocalized charge approximation (QLCA), introduced by Kalman and Golden [15], which was applied to the type-I superlattice [16] and bilayer [15(b), 17] configurations. These latter studies predict that strong Coulomb interactions bring about substantial modifications in the RPA description of plasmon dispersion, most notably, the occurrence of a $q \to 0$ finite-frequency (energy gap) in the out-of-phase longitudinal collective mode. This mode, which is gapless in the RPA, is shown to exhibit this remarkable effect once strong *interlayer* correlations are taken into account. This prediction has recently been convincingly confirmed by the molecular dynamics (MD) simulations of classical charged particle bilayers carried out by Donko *et al* [18] and by Ranganathan and Johnson [19]. On the theoretical side, a recent sum-rule analysis [20] of the long-wavelength behavior of the in-phase and out-of-phase dynamical structure functions further suggests that the energy gap persists over the entire classical to quantum domain all the way down to $T = 0$.

In the low-temperature quantum (Q) domain, three somewhat related theoretical approaches have been used for the calculation of the dielectric matrix and plasmon dispersion primarily in strongly correlated charged-particle bilayers: The approach [Q(i)] followed by Neilson *et al* [21(a)] completely neglects interlayer correlations beyond the RPA under the assumption that the mutual interaction of the layers can be taken into account through the average RPA field. The intralayer correlations are accounted for via a scalar static local field $G(\mathbf{q})$ taken from the Ref. [22] quantum Monte Carlo (QMC) pair correlation function data for the 2D electron liquid in the ground state. The approach [Q(ii)] followed by Swierkowski *et al* [21(b)], by Liu *et al* [21(c)], and by Zhang [23] allows also for interlayer correlations by using a static local field matrix which can be generated by adaptation of the conventional STLS-type approximation scheme [13] to the charged-particle bilayer. Predicated on the assumption that the interlayer interactions are weak, the treatments of Refs. [21(b)] and [21(c)] then use the QMC 2D pair correlation function data of Ref. [22] as an input to the explicit calculations of $G_{11}(\mathbf{q})$ and $G_{12}(\mathbf{q})$. A somewhat more sophisticated approach [Q(iii)] followed by Tanatar and Davoudi [24] is based on an STLS-like approximation scheme [25], referred to as the "qSTLS", which is formulated from a quantum mechanical Wigner function kinetic equation formalism, and which features *dynamical* local field corrections $G_{11}(\mathbf{q},\omega)$ and $G_{12}(\mathbf{q},\omega)$.

The above three approaches share the following features:

- They lack a consistent treatment of strong interlayer correlations: while the intralayer $G_{11}(\mathbf{q})$ is obtained from QMC simulations, the interlayer $G_{12}(\mathbf{q})$ is either taken to be zero or is calculated as a weak perturbation.

- Calculations of the plasmon dispersion are limited to the $r_s < 8$ coupling domain.

- The predicted out-of-phase plasmon dispersion is always acoustic. The mode is softened by exchange-correlation effects such that, for $r_s$ above some critical value $\sim 8$ [21(a), 21(b)], or for sufficiently small layer separations [21(c), 24], the acoustic plasmon ultimately merges with the single-particle excitation region and is quenched by Landau damping.

- The STLS and qSTLS approaches fail to comply with the third-frequency-moment ($<\omega^3>$) sum rule [20, 26], which is known to play a central role in the long-wavelength plasmon dispersion in strongly coupled isolated 2D charged-particle layers [27, 28]. In fact, they completely fail to reproduce in the out-of-phase $<\omega^3>$-sum rule coefficient the crucially important non-vanishing long-wavelength $(q \to 0)$ term.

The approach presented in this paper removes the limitations of the approaches cited above: (i) $G_{11}(\mathbf{q})$ and $G_{12}(\mathbf{q})$ are treated on an equal footing made possible by the availability of pair correlation function data generated by the DMC studies of Rapisarda and Senatore [29]. (ii) The present study is carried out over a wider range of intralayer coupling values extending up to $r_s = 30$, with emphasis placed on understanding how the longitudinal collective mode dispersion is modified by strong interlayer interactions well beyond the RPA. (iii) The method followed in this paper generates a dielectric matrix that almost exactly satisfies the third-frequency-moment sum rule. (iv) It demonstrates that a remarkable $q \to 0$ finite-frequency (energy gap), predicted by Golden, Kalman, and co-workers for strongly coupled layered charged-particle systems in the classical domain [15-20], also exists in the zero-temperature quantum domain. The theoretical confirmation of the existence of the energy gap in degenerate electronic bilayer liquids is the single-most important result of the present work.

The approach followed in this paper is based on the quasilocalized charge approximation (QLCA), an approximation method that has proved to be consistently successful in the description of collective mode dispersion in strongly coupled classical Coulomb liquids, as borne out by comparisons with a series of MD simulations [15(b), 18, 30-32]. The development to be followed here parallels that of a recent companion paper [33] by three of the authors where they extended the QLCA dielectric response function in a way that makes it suitable for the description of collective mode dispersion in strongly coupled 2D Coulomb liquids in the quantum domain. The further extension to the quantum bilayer configuration generalizes the Ref. [33] single-layer scalar formalism to a matrix formalism.

To briefly reiterate what was stated in Ref. [33], the QLCA was formulated by Kalman and Golden some time ago [15(a)] for the express purpose of describing collective mode dispersion in a variety of classical Coulomb liquid configurations [15, 16, 17, 28, 34] in the strong coupling regime. The basis of the formal development of the QLCA is that the dominating feature of the physical state of the plasma with $\Gamma = e^2/(\varepsilon_s a k_B T) \gg 1$ is the quasi-localization of the charges. This physical picture suggests a microscopic equation-of-motion model where the particles are trapped in local potential fluctuations. The particles occupy randomly located (but certainly not uncorrelated) sites and undergo oscillations around them. At the same time, however, the site positions also change and a continuous rearrangement of the underlying quasi-equilibrium configuration takes place. Inherent in the QLC model is the assumption that the two time scales are well separated and that for the description of the rapid oscillating motion, the time average (converted into ensemble average) of the drifting quasi-equilibrium configuration is sufficient.

A unique feature of the extended QLCA developed in this paper is that it reproduces the crucially important exchange-correlation contributions to the $<\omega^3>-$ sum rule coefficients. The connection between the in-phase plasmon dispersion and the in-phase $<\omega^3>$ sum rule coefficient can be inferred from the discussion surrounding the 2D Eqs. (2) and (3) in Ref. [33]. However, the connection between the out-of-phase plasmon dispersion and the out-of-phase sum rule coefficient is even more compelling, since it is the interlayer correlational contribution to the latter that dominates in the $q \to 0$ limit, and it is precisely this contribution that emerges as the $q \to 0$ energy gap [see eq. (19) below]. It is the complete violation of this sum rule by the other competing theories that is responsible for the incorrectly predicted out-of-phase acoustic dispersion in the strong interlayer coupling regime.

Crucial to the extended QLCA approach is the description of the positions of the localized particles in terms of equilibrium pair correlation functions. These latter for the symmetric electronic bilayer liquid have been generated by Rapisarda and Senatore [29(a), 29(b), 29(c)] over a wide range of $r_s$ and $d/a$ values from diffusion Monte Carlo (DMC) simulations. Some of these pair correlation function data have already been published along with the phase diagram [29(a), 29(b), 29(c), 29(d)]. Further results of these simulations, essential to the development of the present analysis, are provided here for the first time. A full compilation of the details of the DMC study is available [29(e)].

Both the DMC simulations and the extended QLCA calculations in the present work are limited to the extent that tunneling between the two layers is ignored. Consequently, the range of validity for the present analysis is necessarily restricted to layer separations $d > a_B$. Note, however, that for $r_s > 1$, this condition still substantially allows for $d < a$, and, consequently, *strong* interlayer interactions.

To summarize, the primary goal of this paper is to construct and analyze the dielectric matrix, $\varepsilon_{AB}(\mathbf{q},\omega)$, for a zero-temperature symmetric bilayer in the strong coupling regime. From this dielectric matrix, we derive the dispersion relation for the longitudinal in-phase and out-of-phase collective modes. The energy gap mentioned above emerges as an intrinsic feature of the out-of-phase mode.

The organization of the paper is as follows: In Section II, we develop the extended QLCA for bilayer systems, we calculate $\varepsilon_{AB}(\mathbf{q},\omega)$ and its diagonalized form with elements $\varepsilon_{\pm}(\mathbf{q},\omega)$. The dielectric matrix contains intralayer exchange-correlation and interlayer correlation contributions beyond the RPA that are ultimately expressed in terms of the pair correlation functions generated from the DMC computer simulations [29] cited above. In Section II, we also analyze the behavior of the $\varepsilon_{\pm}(q,\omega)$ first as functions of $q$ in the static ($\omega=0$) limit and, then in the dynamical ($\omega \neq 0$) domain, as functions of $\omega$ for fixed values of $q$. In Section III, we analyze the plasmon dispersion relations from the zeros of $\varepsilon_{\pm}(\mathbf{q},\omega)$. Conclusions are drawn in Section IV.

## II. DIELECTRIC MATRIX

In this Section, we formulate the longitudinal dielectric response matrix for the strongly coupled charged-particle bilayer liquids at zero temperature. Paralleling the Ref. [33] development, the starting point for the present derivation is the classical (cl) dielectric matrix that results from the QLCA [15(b), (c)]:

$$\varepsilon_{AB}(\mathbf{q},\omega)\big|_{cl} = \delta_{AB} - \sum_{C} \phi_{AC}(q) \frac{nq^2}{m\omega^2}\left[\mathbf{I} - \frac{nq^2}{m\omega^2}\mathbf{D}(\mathbf{q})\right]^{-1}_{CB}; \quad (A,B,C=1,2) \quad (2)$$

$\mathbf{I}$ is the $(2\times 2)$ identity matrix and $(nq^2/m)\mathbf{D}(\mathbf{q})$ is the purely correlational part of the dynamical matrix. Eq. (2) is derived from the microscopic equation-of-motion for the collective coordinates $\vec{\xi}_i^{\mathbf{q}}(t)$, defined through the Fourier representation $\vec{\xi}_A^i(t) = (1/\sqrt{Nm})\sum_{\mathbf{q}} \vec{\xi}_A^{\mathbf{q}}(t)\exp(i\mathbf{q}\cdot\mathbf{x}_A^i)$ relating $\vec{\xi}_A^{\mathbf{q}}$ to the displacement $\vec{\xi}_A^i$ of particle $i$ in layer $A$. The QLCA matrix elements are expressed in terms of the 11, 12 static structure functions, or equivalently, in terms of the 11, 12 pair correlation functions $h_{AB}(r) = (1/N)\sum_{\mathbf{q}}\left[S_{AB}(\mathbf{q}) - \delta_{AB}\right]\exp(i\mathbf{q}\cdot\mathbf{r})$. One obtains [15(b), 17(a)]

$$D_{11}(\mathbf{q}) = \frac{1}{N}\sum_{\mathbf{q}'}\frac{(\mathbf{q}\cdot\mathbf{q}')^2}{q^4}\phi_{11}(q')\left[S_{11}(\mathbf{q}-\mathbf{q}') - S_{11}(\mathbf{q}')\right]$$

$$-\frac{1}{N}\sum_{\mathbf{q}'}\frac{(\mathbf{q}\cdot\mathbf{q}')^2}{q^4}\phi_{12}(q')S_{12}(\mathbf{q}') \quad (3a)$$

$$= \frac{\pi e^2}{q^2}\int_0^{\infty}dr\,\frac{1}{r^2}h_{11}(r)\left[1-4J_0(qr)+6\frac{J_1(qr)}{qr}\right]$$

$$+\frac{\pi e^2}{q^2}\int_0^{\infty}dr\,\frac{rh_{12}(r)}{(r^2+d^2)^{3/2}}\left[1-\frac{3d^2}{r^2+d^2}\right], \quad (3b)$$

$$D_{12}(\mathbf{q}) = \frac{1}{N} \sum_{\mathbf{q}'} \frac{(\mathbf{q} \cdot \mathbf{q}')^2}{q^4} \phi_{12}(q') S_{12}(\mathbf{q} - \mathbf{q}') \tag{4a}$$

$$= -\frac{\pi e^2}{q^2} \int_0^\infty dr \frac{r h_{12}(r)}{(r^2 + d^2)^{3/2}} \left[ 1 - \frac{3d^2}{r^2 + d^2} \right]$$

$$+ \frac{\pi e^2}{q^2} \int_0^\infty dr \frac{r h_{12}(r)}{(r^2 + d^2)^{3/2}} \left[ 1 - 4 J_0(qr) + 6 \frac{J_1(qr)}{qr} \right]$$

$$- \frac{3 \pi e^2 d^2}{q^2} \int_0^\infty dr \frac{r h_{12}(r)}{(r^2 + d^2)^{5/2}} \left[ 1 - 2 J_0(qr) + 2 \frac{J_1(qr)}{qr} \right] \tag{4b}$$

In the long-wavelength $(q \to 0)$ limit, (3a) and (4a) simplify to

$$D_{11}(q \to 0) = \frac{5}{16} \frac{1}{N} \sum_{\mathbf{q}'} \phi_{11}(q') [S_{11}(\mathbf{q}') - 1]$$

$$- \frac{1}{2 N q^2} \sum_{\mathbf{q}'} q'^2 \phi_{12}(q') S_{12}(\mathbf{q}') \tag{5}$$

$$D_{12}(q \to 0) = \frac{1}{2 N q^2} \sum_{\mathbf{q}'} q'^2 \phi_{12}(q') S_{12}(\mathbf{q}')$$

$$+ \frac{5}{16} \frac{1}{N} \sum_{\mathbf{q}'} \phi_{12}(q') S_{12}(\mathbf{q}') \left[ 1 - \frac{11}{5} q'd + \frac{3}{5} (q'd)^2 \right]. \tag{6}$$

The derivation of Eq. (2) is predicated on the assumption that random motions are negligible: this is a reasonable assumption for a low-temperature classical charged-particle bilayer in the strong coupling regime where the potential energy dominates over the thermal energy that is responsible for the random motion so that at sufficiently low temperatures, one can neglect the random motion of the particles. In contrast, for a degenerate system, the low temperature does not ensure that the random motion of the particles is negligible, and one should therefore take account of the ground-state kinetic energy of the particles. In order to accomplish this, we observe that in Eq. (2) the $nq^2/(m\omega^2)$ factor is readily identified as the Vlasov density response function corresponding to the momentum distribution function $f(\mathbf{p}) \sim n\delta(\mathbf{p})$. One may therefore assume that for a Fermi distribution of momenta, the appropriate replacement for $nq^2/(m\omega^2)$ is the Lindhard matrix $\chi_{AB}^{(0)}(\mathbf{q},\omega) = \chi_0(\mathbf{q},\omega)\delta_{AB}$, where

$$\chi_0(\mathbf{q},\omega) = \frac{2}{\hbar}\frac{1}{A}\sum_{\mathbf{p}}\frac{f(|\mathbf{p}+(1/2)\mathbf{q}|) - f(|\mathbf{p}-(1/2)\mathbf{q}|)}{\omega + (\hbar/m)\mathbf{p}\cdot\mathbf{q} + i\eta} \tag{7}$$

The resulting dielectric matrix takes the form

$$\varepsilon_{AB}(\mathbf{q},\omega) = \delta_{AB} - \sum_C \phi_{AC}(q)\chi_0(\mathbf{q},\omega)\left[\mathbf{I} - \chi_0(\mathbf{q},\omega)\mathbf{D}(\mathbf{q})\right]_{CB}^{-1} \tag{8}$$

$(A,B,C = 1,2)$. We note that the dielectric matrix and all other physical quantities can be diagonalized by rotating into the space spanned by the in-phase $(+)$ and out-of-phase $(-)$ directions: for the symmetric bilayer, the resulting matrix elements are $\varepsilon_\pm(\mathbf{q},\omega) = \varepsilon_{11}(\mathbf{q},\omega) \pm \varepsilon_{12}(\mathbf{q},\omega)$, $\phi_\pm(q) = \phi_{11}(q) \pm \phi_{12}(q)$, etc. The diagonalization of (8) therefore results in

$$\varepsilon_\pm(\mathbf{q},\omega) = 1 - \frac{\phi_\pm(q)\chi_0(\mathbf{q},\omega)}{1 - \chi_0(\mathbf{q},\omega)\left[D_{11}(\mathbf{q}) \pm D_{12}(\mathbf{q})\right]}. \tag{9a, b}$$

The $\mathbf{D}(\mathbf{q})$ matrix elements in Eqs. (8) and (9a, b) are formally identical to $D_{11}(\mathbf{q})$ and $D_{12}(\mathbf{q})$ in Eqs. (3) and (4), but it should be borne in mind that the $S_{AB}(\mathbf{q})$ and $h_{AB}(r)$ are now the static structure functions and pair correlation functions appropriate for the zero-temperature symmetric electron bilayer liquid and, as such, these latter quantities embody all the exchange-correlation effects. Accordingly, $D_{11}(\mathbf{q})$ and $D_{12}(\mathbf{q})$ are to be calculated from Eqs. (3) and (4) with the input of the diffusion Monte Carlo (DMC) pair correlation function data presented in this work and in Ref. [29].

A central role is played by the in-phase and out-of-phase third-frequency-moment sum rules which, for the bilayer system under discussion, take the form [20, 26]

$$\frac{1}{\pi}\int_{-\infty}^{\infty}d\omega\omega^{3}\,\text{Im}\frac{1}{\varepsilon_{\pm}(\mathbf{q},\omega)}\bigg|_{exact}$$

$$= -\phi_{\pm}(q)\frac{nq^{2}}{m}\left[\frac{nq^{2}}{m}\phi_{\pm}(q)+3\frac{<E_{kin}>}{m}q^{2}+\frac{nq^{2}}{m}\left[D_{11}(\mathbf{q})\pm D_{12}(\mathbf{q})\right]+\left[\frac{\hbar q^{2}}{2m}\right]^{2}\right]; \quad (10a, b)$$

$D_{11}(\mathbf{q})$ and $D_{12}(\mathbf{q})$ were defined below Eqs. (9a, b); $<E_{kin}>$ is the expectation value of the kinetic energy per particle for the interacting system consisting of a non-interacting part (0) and a correlational (c) part. It is required that the $\varepsilon_{\pm}(\mathbf{q},\omega)$ satisfy these sum rules. In fact, we note that the high-frequency expansion of (9a, b) provides the average energy per particle for the non-interacting system, $<E_{kin}>^{0}$, instead of $<E_{kin}>$. The consequences of this discrepancy, as they pertain to the in-phase $q \to 0$ plasmon dispersion, have been analyzed in companion study [33] of the 2D isolated layer. The findings and quantitative estimates therein certainly should be applicable to the present study for $d/a$ sufficiently large, say, $d/a > 1.5$ where interlayer correlations are weak [18(b), 29(a)]: to reiterate what was stated in Ref. [33], since the missing correlational part of the kinetic energy would act to increase the kinetic energy, Eq. (9a) should lead to an *overestimate* of the softening of the in-phase plasmon dispersion brought about by the effect of exchange and correlations residing in the $D_{11}$ and $D_{12}$ matrix elements. The magnitude of the overestimate decreases with increasing $r_{s}$ and in the $r_{s} \to \infty$ limit, where the total kinetic energy ceases to contribute to the small-$q$ dispersion, one should recover the correct in-phase oscillation frequency characteristic of the isolated 2D Wigner crystal [35] for $d/a$ values well above 1.5. For $d/a < 1.5$, the situation is not expected to be qualitatively different, although the importance of the correlational part of the kinetic energy will depend on the layer separation, since the latter affects the extent to which the particles are localized.

In the case of the out-of-phase plasmon dispersion, the missing correlational part of the kinetic energy is not expected to be an issue, since for $d/a < 1.0$ and in the small-$q$ domain of interest, the $O(q^2 a^2)$ kinetic energy and RPA acoustic plasmon oscillation terms in the sum rule coefficient are overwhelmed by the prominent $O(1)$ term proportional to $[D_{11}(\mathbf{q}) - D_{12}(\mathbf{q})]$ that gives rise to the $q \to 0$ energy gap [see Eqs. (19) and (21) below]. This observation leads us to conclude that, in the long-wavelength domain, agreement between the exact sum rule coefficient (10b) and its third-frequency-moment counterpart generated from (9b) is very good indeed.

The matrix elements $D_{11}(\mathbf{q})$, $D_{12}(\mathbf{q})$ are to be calculated from Eqs. (3b) and (4b) with the input of the DMC pair distribution function data for the spin-unpolarized fluid phase [29(a), 29(c), 29(e), 29(f)]. Figures 1–3 show the pair distribution functions $g_{AB}(r) = 1 + h_{AB}(r)$ as functions of $r/a$ for $10 \le r_s \le 30$, and $0.2 \le d/a \le 1.5$. Note that for $d/a = 0.2$, $g_{12}(r)$ exhibits oscillations that are more pronounced than those exhibited by $g_{11}(r)$ [29(a), 29(b)]. As the layer spacing increases, the oscillations in $g_{11}(r)$ intensify at the expense of those in $g_{12}(r)$. At layer separations $d/a \ge 1.5$, $g_{12}(r) \approx 1$, indicative of isolated 2D layer behavior. For a comparison with the corresponding pair correlation function data for the classical bilayer, see Ref. [36].

The various relevant phases of the symmetric electron bilayer have been mapped in Refs. [29(a), 29(d)]. At $r_s = 10$, the bilayer is in the normal (homogeneous) fluid phase for all values of $d/a$. At $r_s = 20$, the normal (homogeneous) fluid is still the stable phase for $d/a < 0.4$; the bilayer then crystallizes for $0.4 < d/a < 1$; thereafter, for $d/a > 1$, the crystal subsequently melts into a fluid phase. This region in the phase diagram is in close proximity to the boundary separating the fully spin-polarized fluid phase from the unpolarized fluid phase. Along this boundary, the DMC-generated ground state energies of these latter two phases are within combined overlapping error bars, precluding the possibility of specifying one or the other phase with any degree of certainty. For $r_s = 30$ and $d/a \leq 0.2$, the bilayer is in the normal (homogeneous) fluid phase or in its immediate vicinity [29(a), 29(b), (29d)].

Consequently, at $r_s = 10$ and 30, the spin-unpolarized $g_{AB}(r)$ data displayed in Figures 1 and 3 are the appropriate inputs into the Eqs. (3b), (4b) formulas for the computation of $D_{11}(\mathbf{q})$ and $D_{12}(\mathbf{q})$. For $r_s = 20$ and $d/a = 0.2$, the appropriate inputs again are the spin-unpolarized $g_{AB}(r)$ data displayed in Figure 2. At this $r_s$ value, for $d/a = 1.0$ and 1.5, the $g_{AB}(r)$ data for the fully spin-polarized and normal fluids are very nearly the same; thus, it makes little difference which of these data are selected as inputs into Eqs. (3b) and (4b): we choose as inputs the spin-unpolarized $g_{AB}(r)$ data for $d/a = 1.0$ and the fully spin-polarized $g_{AB}(r)$ data for $d/a = 1.5$. These data are also displayed in Figure 2.

To facilitate the collective mode analysis that follows below in Section III, we introduce the more convenient dimensionless quantity $G_{AB}(\mathbf{q}) = -\left[\varepsilon_s q/(2\pi e^2)\right] D_{AB}(\mathbf{q})$, which formally is a static local field correction. One should bear in mind, however, that the physical justification for this term is different from that of the conventional static mean field. $G_{11}(\mathbf{q})$ and $G_{12}(\mathbf{q})$ are shown in Figure 4 as functions of dimensionless in-plane wave number $\bar{q} = q/q_F$ ($q_F = \sqrt{2\pi n}$) for different $r_s$ and $d/a$ values. The small-$q$ behavior of $G_{11}$ and $G_{12}$ is given by Eqs. (5) and (6) which stipulate that to lowest order in $q$, $G_{11}(q \to 0) = -G(r_s, d/a)/q$ and $G_{12}(q \to 0) = G(r_s, d/a)/q$, where

$$G(r_s, d/a) = \frac{1}{2} \int_0^\infty dr \frac{r g_{12}(r)}{(r^2 + d^2)^{3/2}} \left[1 - \frac{3d^2}{r^2 + d^2}\right] > 0. \tag{11}$$

At large-$q$, one can readily show from Eqs. (3) and (4) that

$$G_{11}(q \to \infty) = \left[1 - g_{11}(r = 0)\right] + O(1/q), \tag{12}$$

$$G_{12}(q \to \infty) = \left[1 - g_{12}(r = 0)\right] \exp(-qd). \tag{13}$$

In the $d \to \infty$ limit, one recovers from (12) and (13) the 2D Kimball identity [37]

$$\lim_{q \to \infty} \left[1 - G_{11}(\mathbf{q})\right] = g_{11}(r = 0). \tag{14}$$

valid for any static local field correction $G_{11}(\mathbf{q})$ that one may choose to approximate the exact $G_{11}(\mathbf{q}, \omega)$ for all values of $\omega$. For further clarification, see Ref. [33].

Figure 5 shows how the extended QLCA static in-phase and out-of-phase dielectric functions, $\varepsilon_{\pm}(\bar{q}, \bar{\omega}=0)$, vary with $\bar{q}=q/q_F$ for $r_s=20$, $d/a=0.2$. Figures 6–8 show how their dynamical ($\bar{\omega}\neq 0$) counterparts, $\operatorname{Re}\varepsilon_{\pm}(\bar{q}, \bar{\omega})$, vary with $\bar{\omega}=\hbar\omega/\varepsilon_F$ ($\varepsilon_F=\pi n\hbar^2/m$) over a range of fixed $\bar{q}$-values. We can note a number of points of interest in the behavior of $\operatorname{Re}\varepsilon_{\pm}(\mathbf{q},\bar{\omega})$.

Consider first $\operatorname{Re}\varepsilon_{+}(\mathbf{q},\omega)$:

(i) The in-phase isothermal compressibility is negative for sufficiently high coupling in the classical domain [36] and it must be negative as well in the quantum domain [20, 38]. The compressibility sum rule [20] dictates that $\varepsilon_{+}(\bar{q},0)$ should also develop a first-order pole at $\bar{q}=0$:

$$\varepsilon_{+}(\bar{q}\to 0,0) = K_{+}\frac{2\sqrt{2}r_s}{\bar{q}} + O(1); \tag{15}$$

$K_{+}$ is the compressibility expressible in terms of the physical intralayer and interlayer compressibilities [20]. Its value in the present approximation is calculated from Eqs. (5), (6), and (9a) as

$$K_{+}^{-1} = 1 + \frac{5r_s}{8}\frac{E_{11}+E_{12}}{e^2/a} - \frac{5}{16}\frac{m}{\pi\hbar^2}\sum_{\mathbf{q}'}\phi_{12}(q')h_{12}(q')\left[\frac{11}{5}q'd - \frac{3}{5}(q'd)^2\right], \tag{16}$$

where $E_{11} = (n/2)\int d\mathbf{r}\,\phi_{11}(r)h_{11}(r)$ and $E_{12} = (n/2)\int d\mathbf{r}\,\phi_{12}(r)h_{12}(r)$ are the intralayer and interlayer potential energy per particle, respectively. Thus, $K_{+}<0$ in the strong coupling regime of interest here.

**(ii)** On the interval $0 \leq \bar{q} < \bar{q}_0^+$, where $\bar{q}_0^+$, is the location of a second first-order pole, $\varepsilon_+(\bar{q}, \bar{\omega}=0) < 0$ [Figure 5(a)]. On this interval, the $\varepsilon_+(\bar{q}, \bar{\omega}=0)$ curve is an inverted U with maximum lying below the $\bar{q}$ – axis. For $\bar{q} > \bar{q}_0^+$, $\varepsilon_+(\bar{q}, \bar{\omega}=0)$ descends from positive infinity and approaches unity as $\bar{q} \to \infty$ [Figure 5(a)]. We find that $\bar{q}_0^+ = 3.98$ for $r_s = 20$ and $d/a = 0.2$; $\bar{q}_0^+$ ranges from 3.0 to 4.5 as $r_s$ increases from 10 to 30, its dependence on the $d/a$ ratio being very weak. A similar qualitative behavior has been shown to prevail in the classical domain both for 2D and bilayer systems [36(a), 36(c)]. We note that the $\varepsilon_+(\bar{q}, \bar{\omega}=0)$ curve never penetrates the "forbidden" domain $0 \leq \varepsilon_+(\bar{q}, \bar{\omega}=0) < 1$ [39].

**(iii)** The first-order pole that develops at $\bar{q} = q_0^+$ survives as well for $\bar{\omega} \neq 0$ and in the interval $\bar{q}_0^+ \leq \bar{q} \leq \bar{q}_*^+$, the pole moves along the locus $\bar{\omega} = \bar{\omega}_*^+(\bar{q})$; $\bar{q}_*^+$ [$\approx 5.5$ for $r_s = 20, d/a = 0.2$] is the value of $\bar{q}$ where $\bar{\omega}_*^+(\bar{q})$ reaches the right boundary of the pair excitation continuum. More will be said below about $\bar{q}_*^+$.

**(iv)** On the interval $0 \leq \bar{q} < q_0^+$, $\mathrm{Re}\,\varepsilon_+(\bar{q}, \bar{\omega})$ as a function of $\bar{\omega}$ begins inside the RPA pair excitation continuum with a finite negative value at $\bar{\omega} = 0$ and increases monotonically crossing the $\bar{\omega}$ – axis. For $\bar{q}$ below some critical value $\bar{q}_c^+ < \bar{q}_0^+$, this crossing takes place outside the pair excitation continuum [Figure 6(a), (b)] at the in-phase plasmon excitation frequency $\bar{\omega}_+(\bar{q})$, and thereafter approaches unity as $\bar{\omega} \to \infty$ [$\bar{q}_c^+ \sim 3$ for $r_s = 20, d/a = 0.2$; see Figure 6 (b)].

(**v**) On the interval $q_0^+ \leq \bar{q} \leq q_*^+$, $\operatorname{Re}\varepsilon_+(\bar{q},\bar{\omega})$ starts with a positive value at $\bar{\omega}=0$ and approaches infinity as $\bar{\omega}$ approaches $\bar{\omega}_*^+(\bar{q})$, the location of the pole [Figure 6(c)]. For $\bar{\omega} > \bar{\omega}_*^+(\bar{q})$, $\operatorname{Re}\varepsilon_+(\bar{q},\bar{\omega})$ emerges from negative infinity and crosses the $\bar{\omega}$–axis. This crossing occurs always inside the pair excitation continuum and, as such, cannot represent a collective excitation [Figure 6(c)].

(**vi**) In the interval $\bar{q} > \bar{q}_*^+$, $\operatorname{Re}\varepsilon_+(\bar{q},\bar{\omega})$ is always positive, develops a finite positive peak, and then approaches unity as $\bar{\omega} \to \infty$.

Turning now to $\operatorname{Re}\varepsilon_-(\mathbf{q},\omega)$, we find that its topology is more intricate:

(**i**) For sufficiently high coupling, the compressibility sum rule [20] requires that $\varepsilon_-(\bar{q}\to 0, \bar{\omega}=0)$ assume a finite negative value for the classical bilayer [36(a), 36(c)]. Again, it must be negative as well for the quantum bilayer [20(a)]. However, according to the extended QLCA model, $\varepsilon_-(\bar{q}\to 0, \bar{\omega}=0) = 1 + d\bar{q}^2/(Ga^2) > 1$ [Figure 5(b)]. Indeed, this same defect also shows up in the QLCA treatment of the classical bilayer [17].

(**ii**) On the interval $0 \leq \bar{q} \leq \bar{q}_{00}^-$, $\varepsilon_-(\bar{q},\bar{\omega}=0)$ increases from $\varepsilon_-(\bar{q}=0,\bar{\omega}=0)=1$ and develops a first-order pole at $\bar{q}=\bar{q}_{00}^-$ [=1.885 for $r_s=20, d/a=0.2$; see Figure 5(b)]. For the reason stated in (i) above, this pole must be regarded as unphysical.

(**iii**) On the interval $\bar{q}_{00}^- < \bar{q} < \bar{q}_0^-$, $\varepsilon_-(\bar{q},\bar{\omega}=0)$ behaves in a way similar to $\varepsilon_+(\bar{q},\bar{\omega}=0)$ on the interval $0 < \bar{q} < \bar{q}_0^+$: it develops a second first-order pole at $\bar{q}=\bar{q}_0^-$ [=2.795 for $r_s=20, d/a=0.2$; see Figure 5(b)] and takes the form of an inverted U with maximum well below the $\bar{q}$–axis. Thereafter for $\bar{q} > \bar{q}_0^-$, $\varepsilon_-(\bar{q},\bar{\omega}=0)$ descends from positive infinity and approaches unity as $\bar{q} \to \infty$. Again, a similar qualitative behavior has been reported for the classical bilayer [36(a), 36(c)].

(**iv**) The positive value of $\varepsilon_-(\bar{q}, \omega = 0)$ leads to the formation of a first-order pole [Figure 7(a), (b)] at a finite $\bar{\omega} = \bar{\omega}_{**}^-(\bar{q})$ on the interval $0 \leq \bar{q} \leq \bar{q}_{**}^-$, where $\bar{q}_{**}^-$ [~1.57 for $r_s = 20, d/a = 0.2$] is the $\bar{q}$ – value where $\bar{\omega}_{**}^-(\bar{q})$ reaches the left boundary of the pair excitation continuum. The pole moves along the locus of $\bar{\omega}_{**}^-(\bar{q})$. Again, this pole is spurious.

(**v**) The pole that develops at $\bar{q}_{00}^-$ is within the RPA pair excitation continuum and, in contrast to the behavior of $\bar{q}_0^+$ and $\bar{q}_0^-$ (introduced below), it does not survive for $\bar{\omega} > 0$; it rather generates a complex sequence of maxima and minima in $\text{Re}\,\varepsilon_-(\bar{q}, \bar{\omega})$ in its immediate vicinity. The appearance of the pole represented by $\bar{\omega}_{**}^-(\bar{q})$ above the left boundary of the continuum is, however, intrinsically linked to the existence of the pole at $\bar{q}_{00}^-$ and the former can be regarded as the continuation of the latter outside the continuum.

(**vi**) The first-order pole that develops at $\bar{q} = \bar{q}_0^-$ survives as well for $\bar{\omega} \neq 0$, and in the interval $\bar{q}_0^- < \bar{q} < \bar{q}_*^-$, it moves along the locus $\bar{\omega} = \bar{\omega}_*^-(\bar{q})$; $\bar{q}_*^-$ [$\approx 4$ for $r_s = 20, d/a = 0.2$] is the value of $\bar{q}$ where $\bar{\omega}_*^-(\bar{q})$ reaches the right boundary of the pair excitation continuum. More will be said below about $\bar{q}_*^-$ along with $\bar{q}_*^+$.

, (**vii**) On the interval $0 \leq \bar{q} < \bar{q}_0^-$, the low-frequency behavior of $\text{Re}\,\varepsilon_-(\bar{q}, \bar{\omega})$ is complicated due to the presence of the poles along $\bar{\omega}_{**}^-(\bar{q})$ and in the vicinity of $\bar{q}_{00}^-$. Since both of these poles are spurious, we do not dwell on the details of this behavior.

Within this interval, for $\bar{q}$ below some critical value $\bar{q}_c^-$ [= 2.44 for $r_s = 20, d/a = 0.2$], $\text{Re}\,\varepsilon_-(\bar{q},\bar{\omega})$ crosses the $\bar{\omega}$-axis from below at the out-of-phase plasmon frequency, $\bar{\omega}_-(\bar{q})$, and thereafter approaches unity as $\bar{\omega} \to \infty$ [Figure 7(a), (b)].

(**viii**) On the interval $\bar{q}_0^- \leq \bar{q} \leq \bar{q}_*^-$, $\text{Re}\,\varepsilon_-(\bar{q},\bar{\omega})$ starts with a positive value at $\bar{\omega} = 0$ and approaches infinity as $\bar{\omega} \to \bar{\omega}_*^-(\bar{q})$, the location of the pole [Figure 8(a), (b)]. For $\bar{\omega} > \bar{\omega}_*^-(\bar{q})$, $\text{Re}\,\varepsilon_-(\bar{q},\bar{\omega})$ emerges from negative infinity and crosses the $\bar{\omega}$-axis. This crossing occurs always inside the pair excitation continuum and, as such, cannot represent a collective excitation [Figure 8(a), (b)].

(**ix**) On the interval $\bar{q} > \bar{q}_*^-$, $\text{Re}\,\varepsilon_-(\bar{q},\bar{\omega})$ is always positive, develops a finite positive peak, and then approaches unity as $\bar{\omega} \to \infty$ [Figure 8(c)].

As stated above, the $\bar{q}_0^\pm$ poles survive for $\omega \neq 0$ in the region below and including the right boundary of the RPA pair excitation continuum where they move along $\bar{\omega}_*^\pm(\bar{q})$ loci. Elaborating on this, we analyze $\varepsilon_\pm(\bar{q},\bar{\omega})$ in that region, where [40, 41]

$$\chi_0(\mathbf{q},\omega) = -\frac{m}{\pi\hbar^2}\left\{1 - \frac{1}{2\bar{q}^2}\left[\sqrt{(\bar{\omega}-\bar{q}^2)^2 - 4\bar{q}^2} + \sqrt{(\bar{\omega}+\bar{q}^2)^2 - 4\bar{q}^2}\right]\right\}. \quad (17)$$

From (17), we observe that $\chi_0(\bar{q},\bar{\omega}=-2\bar{q}+\bar{q}^2)$ is always negative on the right boundary of the pair continuum. Then according to (9a, b), the in-phase and out-of-phase dielectric functions on the right boundary each develop a discontinuity at a certain $\bar{q}$ value, say $\bar{q}_*^\pm = \bar{q}_*^\pm(r_s, d/a)$, where the denominators $1 + \phi_{2D}(\bar{q})\chi_0(\bar{q},-2\bar{q}+\bar{q}^2)\left[G_{11}(\bar{q}) \pm G_{12}(\bar{q})\right]$ vanish. The continuation of these discontinuities as first-order poles into the $0 \leq \bar{\omega} < -2\bar{q} + \bar{q}^2$, $\bar{q} \geq 2$ domain is a consequence of the fact that the expression (17) for $\chi_0(\bar{q},\bar{\omega})$ remains negative throughout that entire domain. For a given pair of ($r_s$, $d/a$) values, the loci $\bar{\omega}_*^\pm(\bar{q})$, $\bar{q}_0^\pm \leq \bar{q} \leq \bar{q}_*^\pm$ of all such poles from the $\bar{q}$-axis up to the right boundary then form the families of in-phase and out-of-phase curves shown in Figures 9 and 10. Evidently, the Fourier components of the in-phase and out-of-phase total charge density perturbations are perfectly screened at these $\bar{q}, \bar{\omega}$ values. Our analysis indicates that the pole of $\varepsilon_+(\bar{q},\bar{\omega})$ persists for $r_s$ values all the way down to $\sqrt{2}/[G_{11}(2)+G_{12}(2)] \approx 2.02$ for arbitrary values of $d/a$. This value compares favorably with the Hartree-Fock $r_s = \pi/\sqrt{2} = 2.22$ prediction [6, 38], and with the QMC $r_s \sim 2.03$ value [22] and experimentally observed value $r_s = 1.71$ [38] for the onset of negative compressibility in 2D degenerate electron liquids. The same critical $r_s \approx 2.02$ value results as well for the out-of-phase pole as well for $d/a \geq 1$. For sufficiently small layer spacings, however, the existence of the out-of-phase pole can become strongly $d/a$-dependent. For example, the out-of-phase pole does not develop at $r_s = 10$ and $d/a = 0.2$ [no $r_s = 10$ curve displayed in Figure 9(b)], whereas it does develop at $d/a = 0.5$ [see Figure 9(a)] indicating that there is some value between 0.2 and 0.5, below which $\varepsilon_-(\bar{q},\bar{\omega})$ ceases to develop any pole behavior.

### III. PLASMON DISPERSION

We turn now to the calculation of plasmon dispersion in the strongly coupled symmetric charged-particle bilayer liquid. We use the formulations of Stern [40] and Isihara [41] for the zero-temperature $\chi_0(\mathbf{q},\omega)$ in the extended QLCA formulas (9a, b) for $\varepsilon_\pm(\mathbf{q},\omega)$. The mode frequencies above and on the left boundary of the pair continuum are then calculated by equating $\varepsilon_\pm(\mathbf{q},\omega)$ to zero with $\chi_0(\mathbf{q},\omega)$ given by [40, 41]

$$\chi_0(\mathbf{q},\omega) = -\frac{m}{\pi\hbar^2}\left\{1 + \frac{1}{2\bar{q}^2}\left[\sqrt{(\bar{\omega}-\bar{q}^2)^2 - 4\bar{q}^2} - \sqrt{(\bar{\omega}+\bar{q}^2)^2 - 4\bar{q}^2}\right]\right\} \quad (18)$$

In the $q \to 0$ limit, $\omega_+(0) = 0$ and

$$\omega_-(\bar{q} \to 0) = \omega_0\sqrt{2aG} \equiv \omega_{GAP}, \quad (19)$$

with $G = G(r_s, d/a)$ given by Eq. (11); $\omega_0 = \sqrt{2\pi n e^2/(ma)}$ is a nominal 2D plasma frequency. The correlation-induced energy gap (19), displayed in Figure 11 as a function of $r_s$ and $d/a$, is a unique feature both of the QLCA approach of Refs. [15] – [17] and of the extended QLCA approach of the present paper. As we have stated above, its existence in classical bilayers has been confirmed by recent MD simulations [18, 19]. Since the physical conditions leading to the finite-frequency gap are similar in the classical and quantum domains, there is little doubt that the results of the classical simulations are relevant to the present work as well. In addition, the sum rule analysis of Ref. [20] provides a further theoretical basis for expecting similar such behavior in classical and quantum bilayers.

At long wavelengths, the in-phase and out-of-phase plasmon frequencies [20]

$$\omega_+^2(q \to 0) = 2\omega_0^2 qa\left[1 - \frac{1}{2}qd + \frac{1}{2}qa\left(\frac{3}{4}r_s\varepsilon_{kin}^0 + \gamma_+\right)\right], \quad (20)$$

$$\omega_-^2(q \to 0) = \omega_{GAP}^2 + \omega_0^2 q^2 ad\left[1 + \frac{a}{d}\left(\frac{3}{4}r_s\varepsilon_{kin}^0 + \gamma_-\right)\right], \quad (21)$$

result from Eqs. (5), (6), (9a, b) and (18); $\varepsilon_{kin}^0 = 1/r_s^2$ is the non-interacting part of the kinetic energy per particle in Rydberg units. The last right-hand-side members of (20) and (21) can be expressed in terms of the static structure functions:

$$\gamma_\pm = \frac{5a}{32}\int_0^\infty dq'\left[S_{11}(q')-1\right] \pm \frac{5a}{32}\int_0^\infty dq'\left[1-\frac{11}{5}q'd+\frac{3}{5}(q'd)^2\right]S_{12}(q')\exp(-q'd). \quad (22\text{a, b})$$

Or, alternatively, in terms of potential energies:

$$\gamma_\pm = \frac{5}{16}\left[\frac{E_{11}\pm E_{12}}{(e^2/a)}\right] \mp \frac{7}{8}\frac{d^2}{a^2}\frac{\omega_{GAP}^2}{\omega_0^2} \mp \frac{33}{16}\frac{d^4}{a}\int_0^\infty dr\frac{rh_{12}(r)}{(r^2+d^2)^{5/2}}; \quad (23\text{a, b})$$

$E_{11} = (n/2)\int d\mathbf{r}\phi_{11}(r)h_{11}(r)$ is the intralayer potential energy per particle and $E_{12} = (n/2)\int d\mathbf{r}\phi_{12}(r)h_{12}(r)$ is the interlayer potential energy per particle. Our calculations indicate that the $\gamma_\pm$ plasmon dispersion coefficients are negative so that they always act to soften the in-phase and out-of-phase plasmon dispersion curves.

The dielectric matrix elements (9a, b) do not take account of collisional (multi-pair excitations) damping, leaving Landau damping as the sole mechanism responsible for the decay of collective excitations in the present study. At zero temperature, the Landau damping is confined to the RPA pair excitation continuum region of the $q, \bar{\omega}$–plane. For $\bar{\omega} \geq 0$, the equations for the left and right boundaries of the continuum region are given by $\bar{\omega} = 2\bar{q} + \bar{q}^2$ and $\bar{\omega} = -2\bar{q} + \bar{q}^2$, respectively.

The straightforward calculation of the plasmon oscillation frequencies in the region $\bar{\omega} \geq 2\bar{q} + \bar{q}^2$, $\bar{q} \geq 0$ is then carried out by substituting the Lindhard density response function (18) into (9a, b) and equating $\varepsilon_\pm(\mathbf{q},\omega)$ to zero. We obtain

$$\overline{\omega}_{\pm}(\overline{q};r_s,d/a) = A_{\pm}\overline{q}\sqrt{\frac{\overline{q}^2}{(A_{\pm}-1)^2} + \frac{4}{(2A_{\pm}-1)}}, \qquad (24a, b)$$

$$A_{\pm} = 1 + \frac{\sqrt{2}r_s}{\overline{q}}\left\{\left[1\pm\exp(-\sqrt{2}\overline{q}\overline{d})\right] - \left[G_{11}(\overline{q}) \pm G_{12}(\overline{q})\right]\right\}, \qquad (25a, b)$$

The analytical formulas (24a, b) provide the plasmon dispersion curves up to the point where they make first contact with the left boundary of the pair continuum at $\overline{q} = \overline{q}_c^{\pm}$. The in-phase and out-of-phase dispersion curves and their RPA counterparts are displayed in Figures 12–15 for $r_s = 10, 20, 30$ and $d/a$ values ranging from 0.2 to 1.5.

Figures 12 and 13 show that the in-phase plasmon mode is not qualitatively different from the similar mode of the isolated 2D layer [33]. In particular, for $q \to 0$, we see from Eq. (20) that the in-phase mode exhibits the typical $\omega \sim \sqrt{q}$ dispersion which is always softened by the $\gamma_+$ dispersion coefficient portraying interlayer correlations and intralayer exchange and correlations.

Figures 14 and 15 show how the energy gap can dramatically modify the acoustic dispersion of the out-of-phase plasmon. With increasing layer spacing and consequently decreasing interlayer correlations, the gap frequency, $\omega_{GAP}$, becomes less and less pronounced and all but disappears for $d/a > 1.5$ at which point the separated layers become practically uncorrelated [29(a), see also Figs. 1(b), 2(b)]; at $d/a = 1.5$, Figure 15(b) shows the dispersion of the out-of-phase plasmon to be very nearly acoustic. This is precisely what was predicted by Kalman *et al* [17] for the classical bilayer and subsequently confirmed by the MD simulations of Donko *et al* [18]. The effect of the single-pair excitations can be assessed from Eq. (20) and from Figures 14 and 15 showing the RPA pair excitation continuum: as long as the layer separation is not too large $(d/a \leq 1)$, the out-of-phase gapped mode lies well above the continuum and is therefore entirely immune to Landau damping. This is in marked contrast to the findings of Neilson *et al* [21] and Tanatar and Davoudi [24]: their approaches predict that the out-of-phase plasmon is acoustic and is softened by exchange-correlations so that it is no longer immune to Landau damping beyond some critical $r_s$ value. On the other hand, this softening of the slope of the dispersion curve is a common feature of all the above theoretical approaches including the QLCA.

We remind the reader of one noteworthy feature of the energy gap in the quantum bilayer: While it is true that *relative to the Fermi energy of the non-interacting 2D electron gas*, the magnitude of the gap *increases* with decreasing carrier density (increasing $r_s$), it is, in fact, the case that the *absolute* gap energy *decreases* with decreasing density according to the formula $\hbar\omega_{GAP} = \left[10.79/r_s^2\right]\bar{\omega}_{GAP}$ (meV) [for GaAs/Ga$_{1-x}$Al$_x$As]. Figure 11(b) illustrates this point. This is in marked contrast to the classical bilayer where the absolute gap frequency (in Hz) *increases* with increasing intralalyer coupling parameter.

As to the region $0 \leq \bar{\omega} \leq -2\bar{q} + \bar{q}^2$, $\bar{q} \geq 2$ on or below the right boundary of the pair continuum, our calculations indicate that the determinant of the dielectric matrix does not possess any zeros there. Consequently, there are no collective excitations in this region.

## IV   CONCLUSIONS

In this paper, we have developed and analyzed a dielectric matrix for strongly coupled symmetric charged-particle bilayers at zero-temperature. This has been carried out over a range of coupling values $10 \leq r_s \leq 30$ not addressed in other competing theories. Our analysis is based on an extension of the classical quasi-localized charge approximation (QLCA) [15] into the quantum domain. The development of the extended QLCA matrix formalism of the present work parallels the development of the Ref. [33] scalar formalism for the isolated 2D layer.

The extended QLCA formalism of the present work, like its classical counterpart, requires the input of the intralayer and interlayer pair distribution functions. In fact, the resulting plasmon dispersion calculations are quite sensitive to the structure of the interlayer correlations and therefore the precise determination of the latter is essential. Pair correlation function data, generated from diffusion Monte Carlo simulations and displayed in Figures 1, 2, 3, are used in the present calculations.

The calculation of the dielectric matrix results in explicit expressions for the in-phase and out-of-phase dielectric response functions, $\varepsilon_+(\mathbf{q},\omega)$ and $\varepsilon_-(\mathbf{q},\omega)$, respectively, leading to a description of the two longitudinal collective modes.

The Eq. (9a) in-phase dielectric response function, $\varepsilon_+(\mathbf{q},\omega)$, exactly satisfies its third-frequency-moment sum rule in the $r_s \to \infty$ limit, thereby guaranteeing recovery of the correct 2D plasmon dispersion at long wavelengths in the $d \to \infty$ isolated 2D layer limit [33, 35]. More importantly, the dominance at long wavelengths of the energy gap contribution (11) to the Eq. (9b) out-of-phase dielectric response function, $\varepsilon_-(\mathbf{q},\omega)$, all but guarantees near-perfect satisfaction of the out-of-phase third-frequency-moment sum rule for *arbitrary* $r_s$ values. That is to say, in the small-$q$ domain, the correlational part of the kinetic energy that is missing from Eq. (9b) is of little consequence, since it is absolutely overwhelmed by the energy gap contribution.

The main result of the present work is the demonstration of the existence of the long-wavelength finite-frequency energy gap (19) [Figure 11(b)] in the out-of-phase plasmon dispersion in the zero-temperature quantum domain. The existence of the energy gap in classical layered charged-particle systems has already been predicted and extensively analyzed over the past decade in a series of theoretical works [15-18, 20]. Recent molecular dynamics simulations [18, 19] now confirm its existence in classical charged-particle bilayer liquids over a wide range of intralayer coupling strengths and for interlayer spacing $d < 1.5a$. By contrast, the more traditional STLS and qSTLS approaches [21, 24] predict that the out-of-phase plasmon is an acoustic excitation which should ultimately merge with the pair continuum when $r_s$ exceeds some critical coupling value [21]. In the present work we find that as long as the layer separation is sufficiently small ($d < 1.5a$), the presence of the energy gap ensures that the out-of-phase plasmon is always well above the continuum and is thus immune to Landau damping.

We recall that in the present work and in the companion DMC simulations [29], tunneling between the two layers is ruled out so that the range of validity of the extended QLCA is necessarily restricted to layer separations $d > a_B$. We call attention to the marked distinction between the energy gap reported in the present paper and the $q = 0$ plasmon gap in the out-of-phase mode reported by Das Sarma and Hwang [42]: the former is brought about solely by strong interlayer correlations in the absence of interlayer quantum tunneling, while the latter is brought about solely by interlayer tunneling in the absence of particle correlations.

As to experimental verification in the quantum domain, the existing observations on semiconductor electronic bilayers at small-$r_s$ and high-$q$ values [43] can be reconciled with the miniscule energy gap that would exist in this parameter range. The ultimate verification of the existence of the energy gap in the zero-temperature quantum domain awaits inelastic light scattering experiments on high-$r_s$ multiple quantum well structures.

**ACKNOWLEDGMENTS**

This material is based upon work supported by the National Science Foundation under Grant Nos. PHY-0206554 and PHY-0206695. Hania Mahassen's participation in this research was also partially sponsored by the Vermont Space Grant Consortium and by NASA under grant number NGT5-40110.

* Also at Department of Physics, University of Vermont, Burlington, Vermont 05405, USA

† Present Address: Product and Business Development Group, Banca IMI, Milano, Italy.

**FIGURE CAPTIONS**

FIGURE 1:

Diffusion Monte Carlo intralayer (11) and interlayer (12) pair distribution functions, $g_{ij}(r)$, for a symmetric electronic bilayer in the normal fluid phase at $r_s = 10$ and $d/a = 0.2, 0.5, 1.0, 1.5$. (a) $g_{11}(r)$: the curve with the highest peak corresponds to $d/a = 1.5$; (b) $g_{12}(r)$: the curve with the highest peak corresponds to $d/a = 0.2$.

FIGURE 2:

Diffusion Monte Carlo intralayer (11) and interlayer (12) pair distribution functions for a symmetric electron bilayer at $r_s = 20$ and $d/a = 0.2$ (normal fluid), 1.0 (normal fluid), 1.5 (fully spin-polarized fluid). (a) $g_{11}(r)$: the curve with the highest peak corresponds to $d/a = 1.5$; (b) $g_{12}(r)$: the curve with the highest peak corresponds to $d/a = 0.2$.

FIGURE 3:

Diffusion Monte Carlo intralayer (11) and interlayer (12) pair distribution functions for a symmetric electron bilayer in the normal fluid phase at $r_s = 30$ and $d/a = 0.2$. The dashed and full curves label $g_{12}(r)$ and $g_{11}(r)$, respectively.

FIGURE 4:

Intralayer (11; solid curves) and interlayer (12: dashed curves) local field factors for the symmetric electronic bilayer as functions of $\bar{q} = q/q_F$; $q_F = \sqrt{2\pi n}$. (a) (normal fluid) $r_s = 10$ and $d/a = 0.2, 0.5, 1.0, 1.5$; the lowest lying $G_{11}(\bar{q})$ curve corresponds to $d/a = 0.2$; the highest lying $G_{12}(\bar{q})$ curve corresponds to $d/a = 0.2$. (b) $r_s = 20$ and $d/a = 0.2$ (normal fluid), 1.0 (normal fluid), and 1.5 (fully spin-polarized fluid); the lowest lying $G_{11}(\bar{q})$ curve corresponds to $d/a = 0.2$; the highest lying $G_{12}(\bar{q})$ curve corresponds to $d/a = 0.2$. (c) (normal fluid) $r_s = 30$ and $d/a = 0.2$.

FIGURE 5:

In-phase $(+)$ and out-of-phase $(-)$ static dielectric functions for the normal fluid phase at $r_s = 20$, $d/a = 0.2$; (a) $\varepsilon_+(\bar{q}, \omega = 0)$ develops a first-order pole at $\bar{q}_0^+ = 3.978$; (b) $\varepsilon_-(\bar{q}, \omega = 0)$ develops first-order poles at $\bar{q}_{00}^- = 1.885$ and $\bar{q}_0^- = 2.795$.

FIGURE 6:

$\operatorname{Re}\varepsilon_+(\bar{q},\bar{\omega})$ vs $\bar{\omega} = \hbar\omega/\varepsilon_F$ for $\bar{q} = 1.0, 3.0, 4.2$: for the normal fluid phase at $r_s = 20$, $d/a = 0.2$; $\varepsilon_F = \pi n \hbar^2/m$.

FIGURE 7:

$\operatorname{Re}\varepsilon_-(\bar{q},\bar{\omega})$ vs $\bar{\omega} = \hbar\omega/\varepsilon_F$ for $\bar{q} = 0.5, 1.0, 2.5$; for the normal; fluid phase at $r_s = 20$, $d/a = 0.2$.

FIGURE 8:

$\operatorname{Re}\varepsilon_-(\bar{q},\bar{\omega})$ vs $\bar{\omega} = \hbar\omega/\varepsilon_F$ for $\bar{q} = 3.0, 4.0, 4.1$; for the normal fluid phase at $r_s = 20$, $d/a = 0.2$.

FIGURE 9:

In the region $0 \leq \bar{\omega} \leq -2\bar{q} + \bar{q}^2, \bar{q} \geq 2$: Loci of first-order poles of the dielectric response functions, $\varepsilon_\pm(\bar{q},\bar{\omega})$, calculated from Eqs. (9a, b), (3b), (4b), and (17) for the normal fluid phase at $d/a = 0.2$; (a) in-phase curves for $r_s = 10, 20, 30$; (b) out-of-phase for $r_s = 20, 30$.

FIGURE 10:

In the region $0 \leq \bar{\omega} \leq -2\bar{q} + \bar{q}^2, \bar{q} \geq 2$: Loci of first-order poles of the dielectric response functions, $\varepsilon_\pm(\bar{q}, \bar{\omega})$, calculated from Eqs. (9a, b), (3b), (4b), and (17) for the normal fluid phase; (a) in-phase and out-of-phase curves for $d/a = 0.5$, $r_s = 10$; (b) in-phase and out-of-phase curves for $d/a = 1.0$, $r_s = 10, 20$.

FIGURE 11:

Energy gap values as a function of layer separation $d/a$ for $r_s = 10$ and 20; (a) in units of the nominal 2D plasma frequency $\omega_0 = \sqrt{2\pi n e^2/(\varepsilon_s m a)}$; (b) in energy units for GaAs/AlGaAs.

FIGURE 12:

In-phase plasmon dispersion curves for the symmetric electronic bilayer: (a) $d/a = 0.2$ and $r_s = 10, 20, 30$; (b) $d/a = 0.5$ and $r_s = 10$. The full curves are calculated from Eqs. (9a), (18) [or equivalently, from (24a), (25a)], (3b), and (4b) with the input of the diffusion Monte Carlo pair distribution function data (shown in Figures 1-3) for the normal fluid. The dashed RPA curves are calculated from Eq. (9a) with $D_{11}(\mathbf{q})$ and $D_{12}(\mathbf{q})$ set equal to zero; in (a), the highest $r_s$ value corresponds to the highest RPA curve. The hachured region is the RPA pair continuum; $\bar{q} = q/q_F, \bar{\omega} = \omega/\omega_F; q_F = \sqrt{2\pi n}, \hbar \omega_F = \varepsilon_F = \pi n \hbar^2/m$.

FIGURE 13:

In-phase plasmon dispersion curves for the symmetric electronic bilayer: (a) $d/a = 1.0$ and $r_s = 10$ (normal fluid), 20 (normal fluid); (b) $d/a = 1.5$ and $r_s = 10$ (normal fluid), 20 (fully spin-polarized fluid). The full curves are calculated from Eqs. (9a), (18) [or equivalently, from (24a), (25a)], (3b), and (4b) with the input of the diffusion Monte Carlo pair distribution function data (shown in Figures 1-3). The dashed RPA curves in are calculated from Eq. (9a) with $D_{11}(\mathbf{q})$ and $D_{12}(\mathbf{q})$ set equal to zero; in (a) and (b), the highest $r_s$ value corresponds to the highest RPA curve. The hachured region is the RPA pair continuum; $\bar{q}=q/q_F, \bar{\omega}=\omega/\omega_F; q_F=\sqrt{2\pi n}, \hbar\omega_F=\varepsilon_F=\pi n\hbar^2/m$.

FIGURE 14:

Out-of-phase plasmon dispersion curves for the symmetric electronic bilayer: (a) $d/a=0.2$ and $r_s=10, 20, 30$; (b) $d/a=0.5$ and $r_s=10$. The full curves are calculated from extended QLCA Eqs. (9b), (18) [or equivalently, from (24b), (25b)], (3b), and (4b) with the input of the diffusion Monte Carlo pair distribution function data for the normal fluid phase (shown in Figures 1-3). The inset in Figure 14(b) shows the crossing of the in-phase and out-of-phase dispersion curves. The dashed RPA acoustic curves are calculated from Eq. (9b) with $D_{11}(\mathbf{q})$ and $D_{12}(\mathbf{q})$ set equal to zero; in (a), the highest $r_s$ value corresponds to the highest lying RPA curve.

FIGURE 15:

Out-of-phase plasmon dispersion curves for the symmetric electronic bilayer for (a) $d/a=1.0$ $r_s=10$ (normal fluid), 20 (normal fluid); (b) $d/a=1.5$, $r_s=10$ (normal fluid), 20 (fully spin-polarized fluid). The full curves are calculated from extended QLCA Eqs. (9b), (18) [or equivalently, from (24b), (25b)], (3b), and (4b) with the input of the diffusion Monte Carlo pair distribution function data (shown in Figures 1-3). The dashed RPA acoustic curves are calculated from Eq. (9b) with $D_{11}(\mathbf{q})$ and $D_{12}(\mathbf{q})$ set equal to zero; in (a) and (b), the highest $r_s$ value corresponds to the highest lying RPA curve.

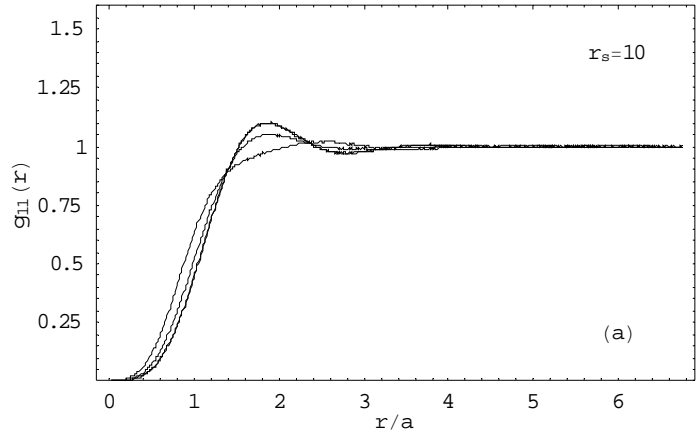
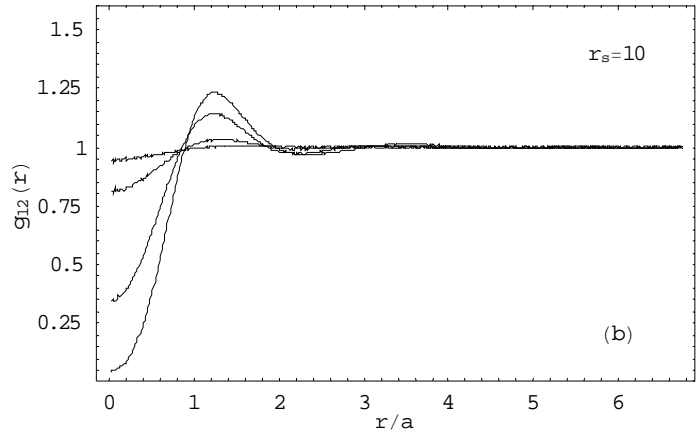

FIGURE 1

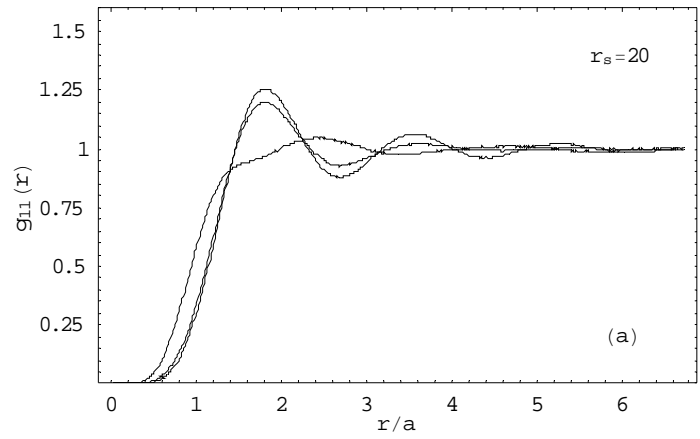

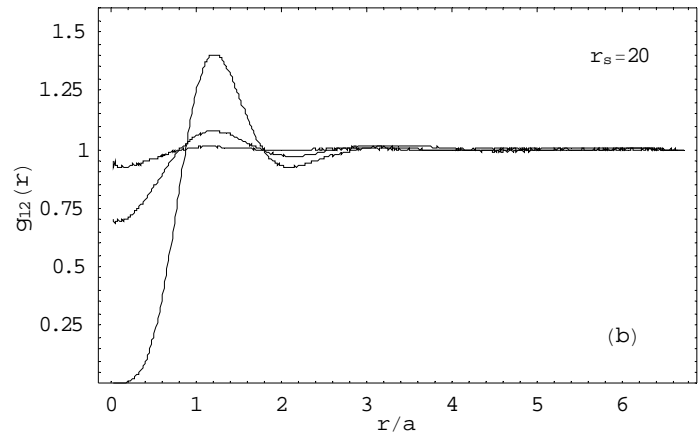

FIGURE 2

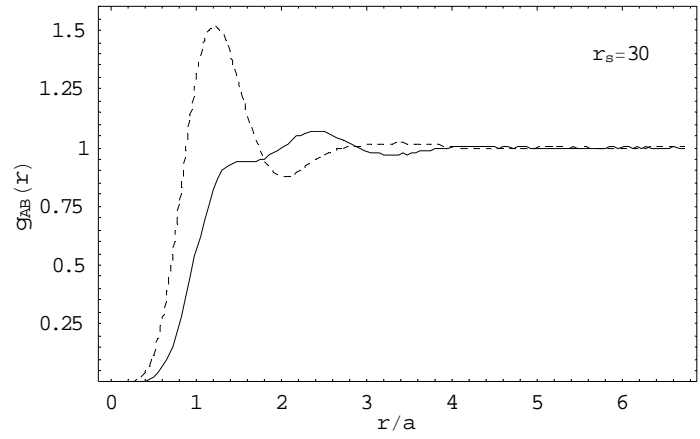

FIGURE 3

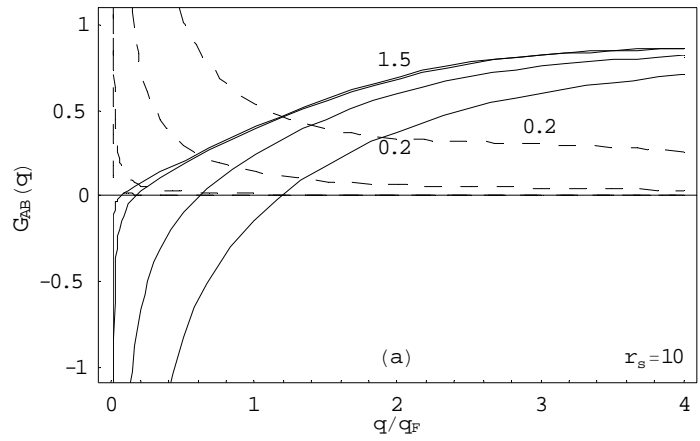

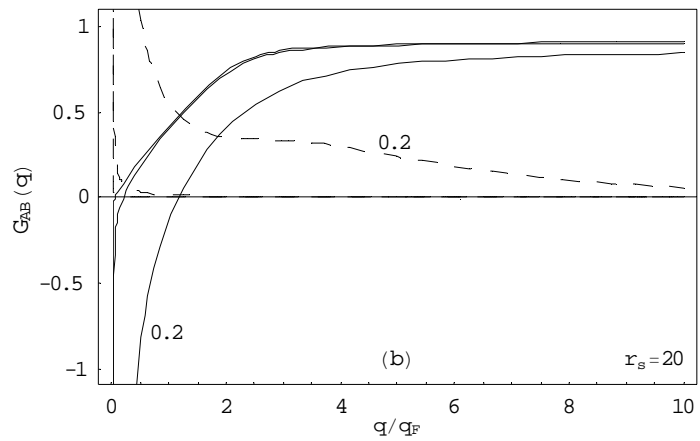

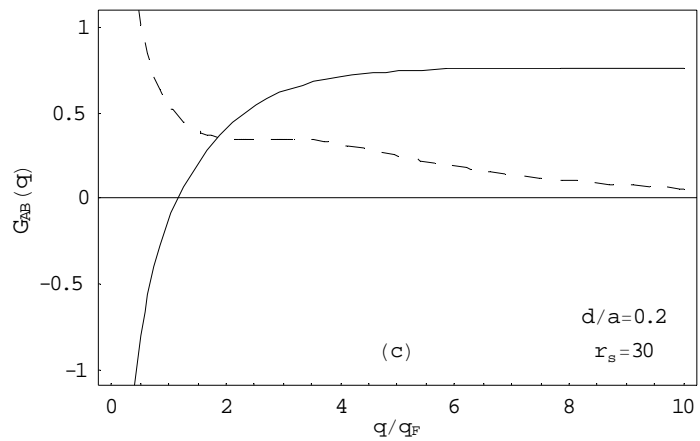

FIGURE 4

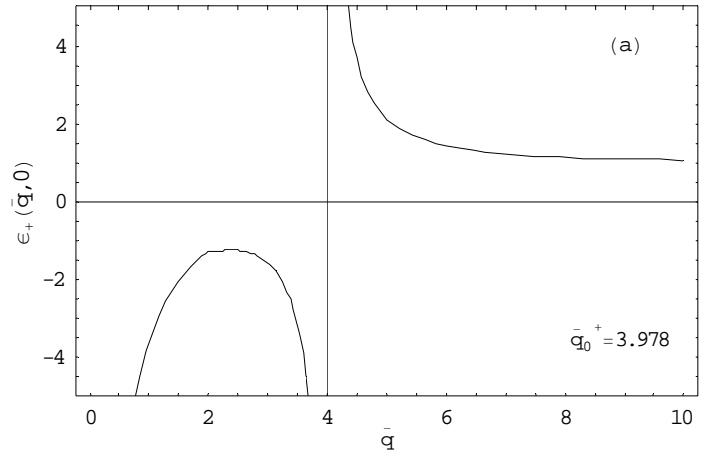

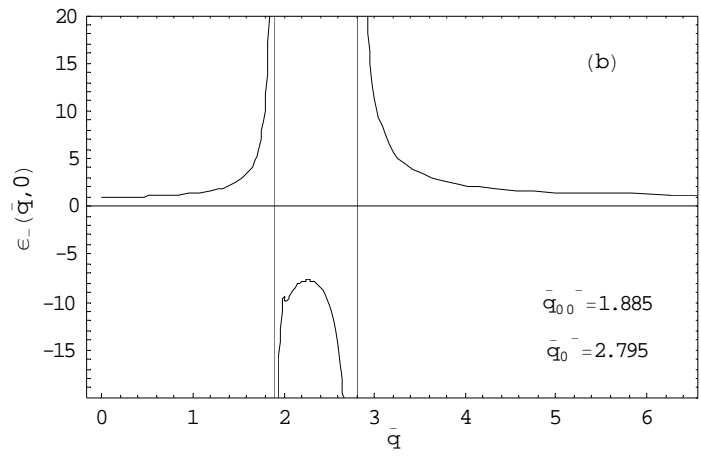

FIGURE 5

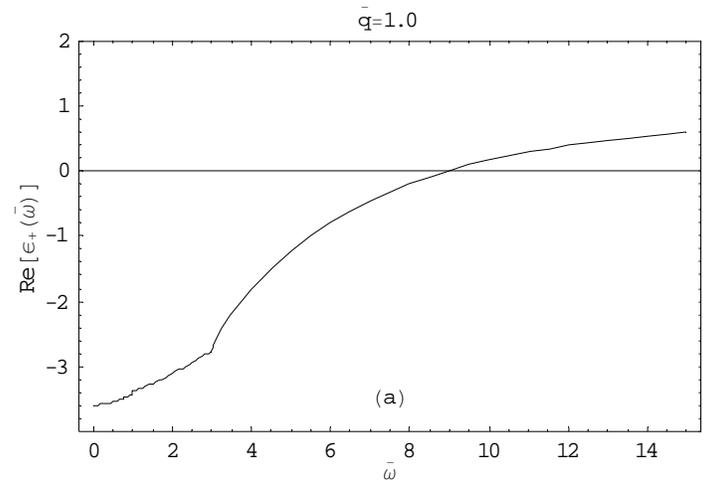

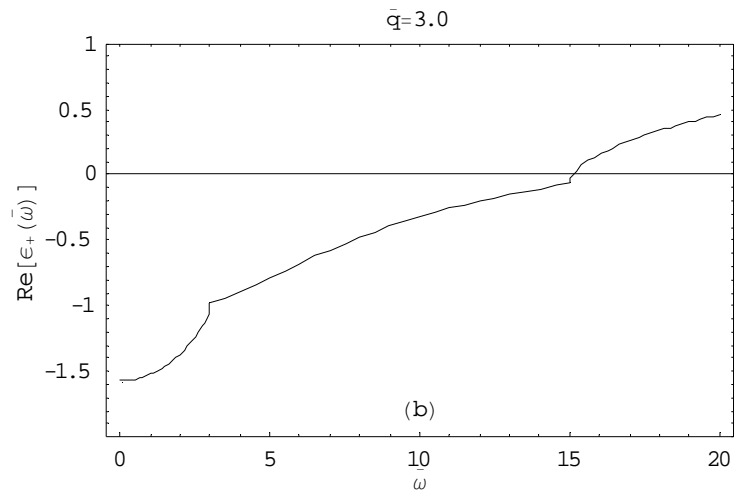

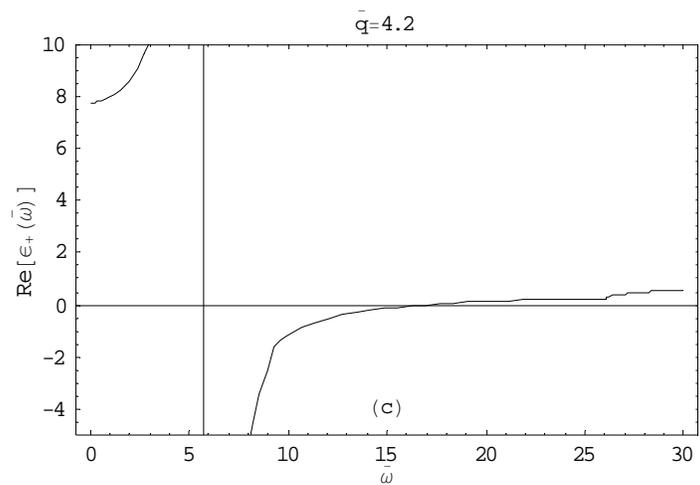

FIGURE 6

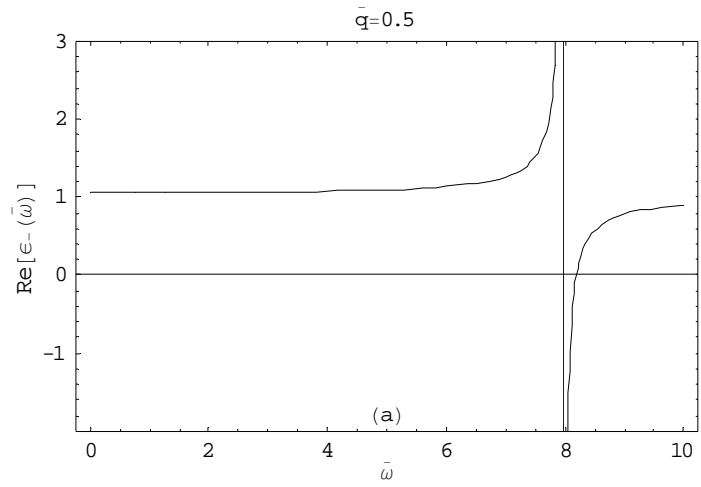

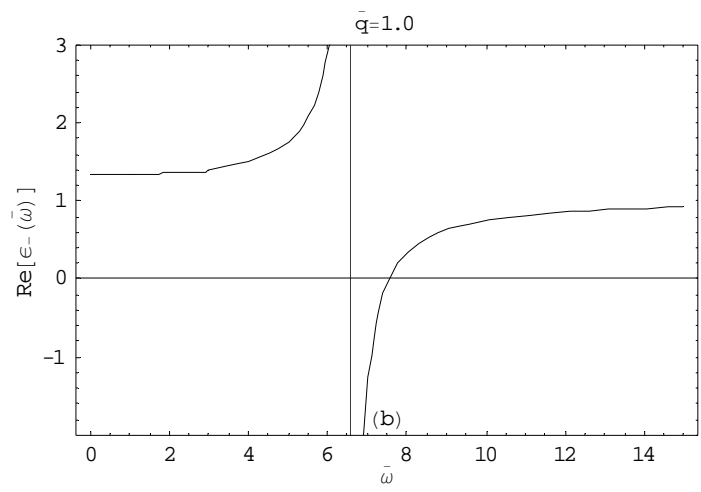

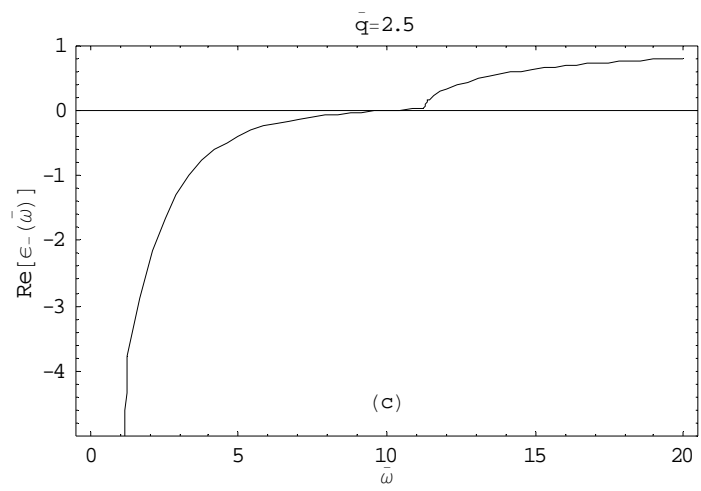

FIGURE 7

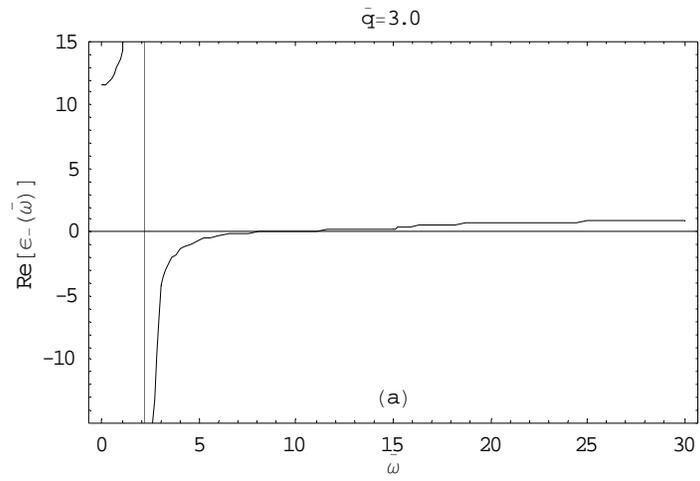

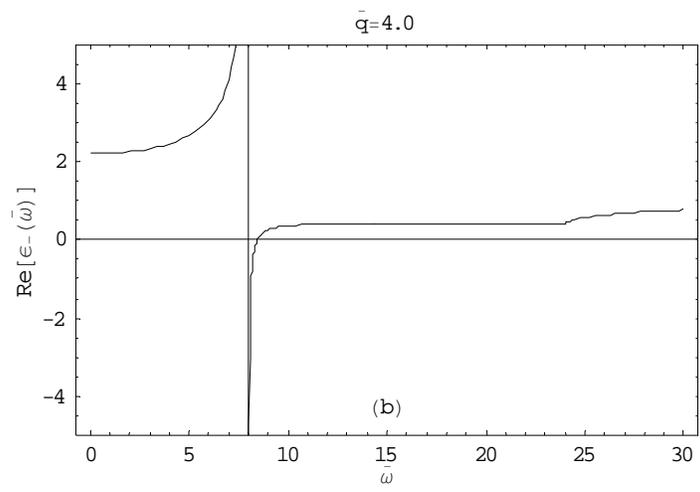

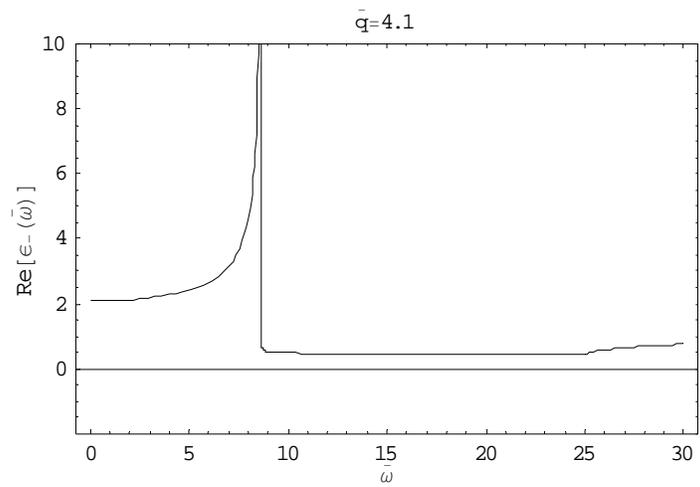

FIGURE 8

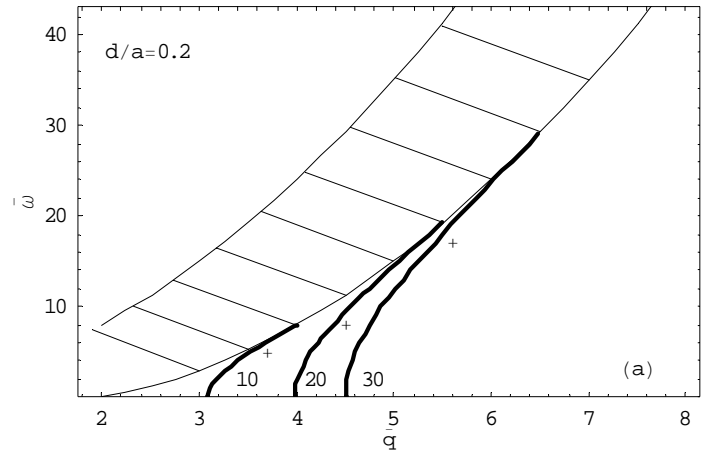

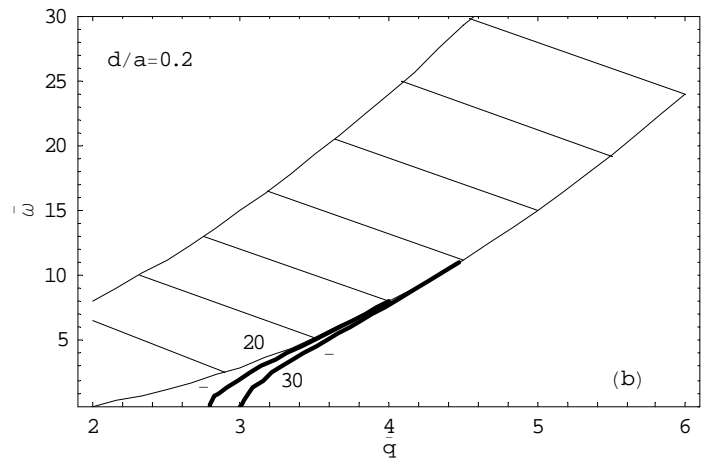

FIGURE 9

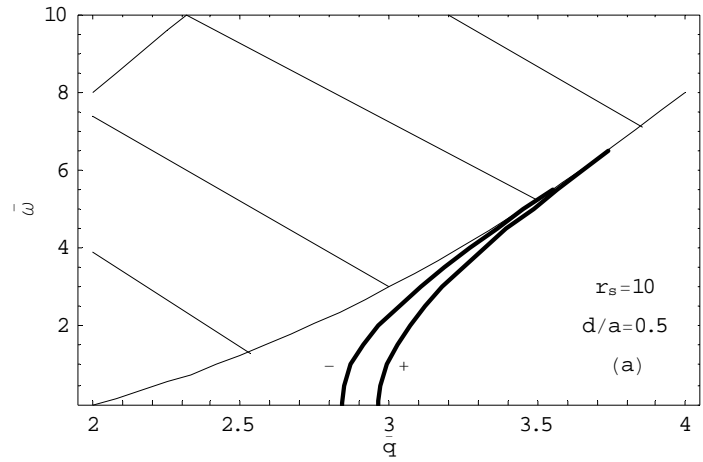

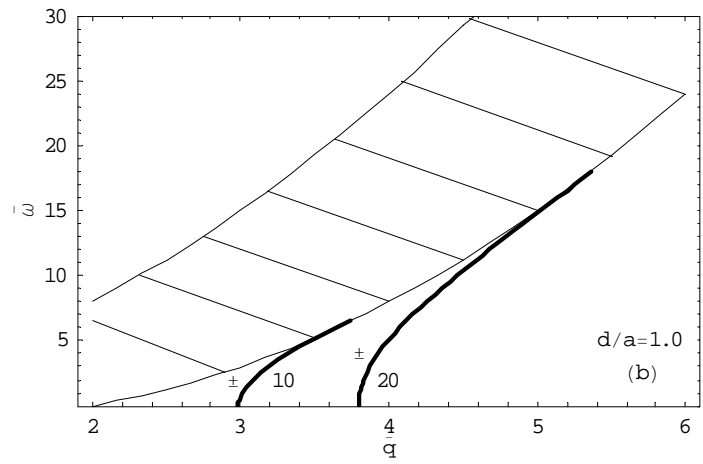

FIGURE 10

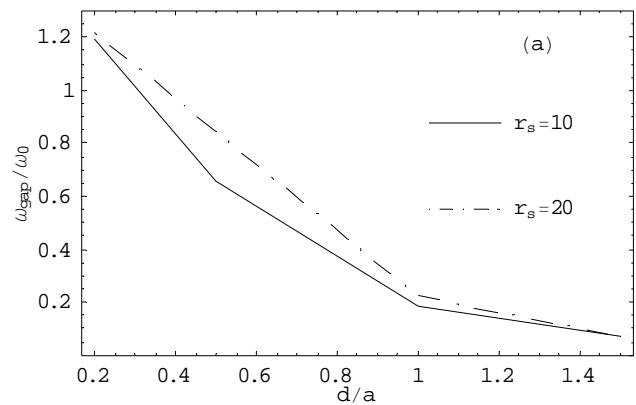

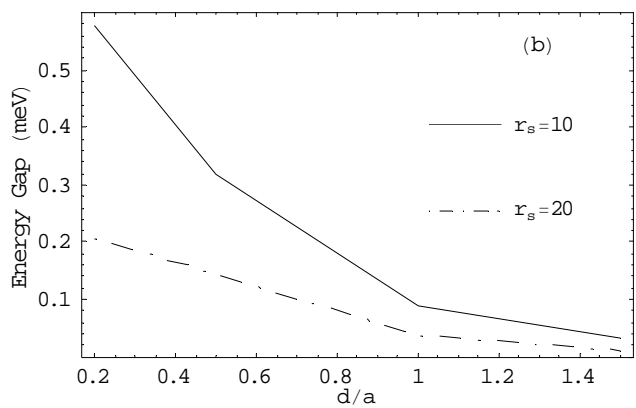

FIGURE 11

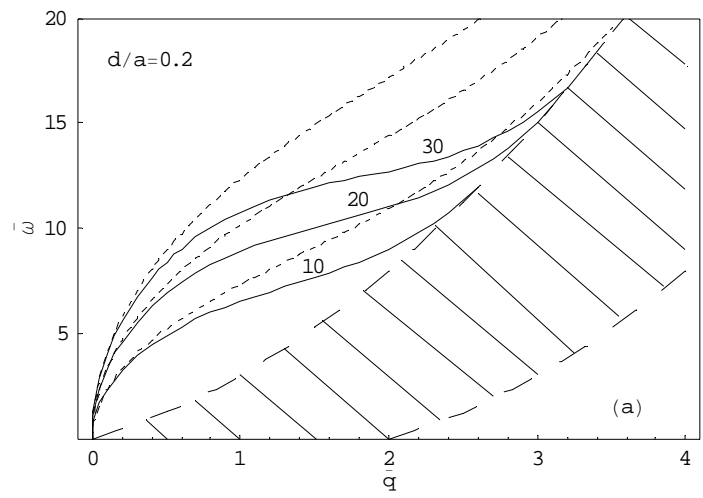

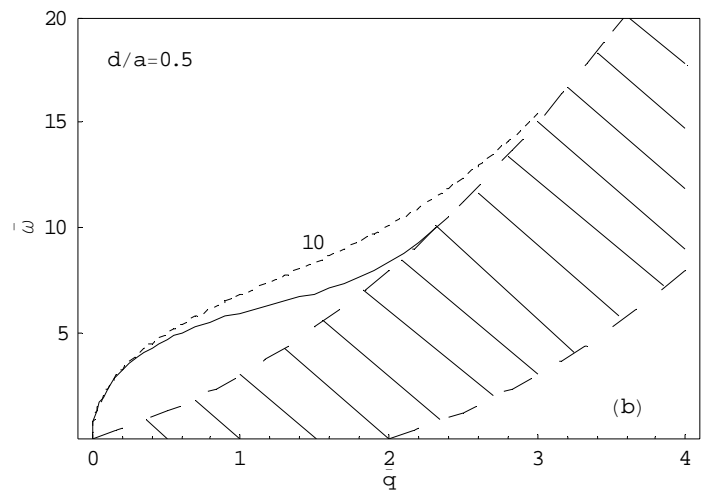

FIGURE 12

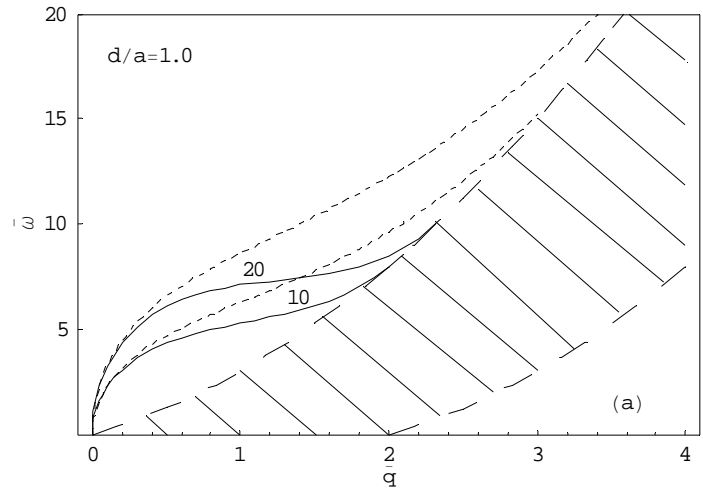

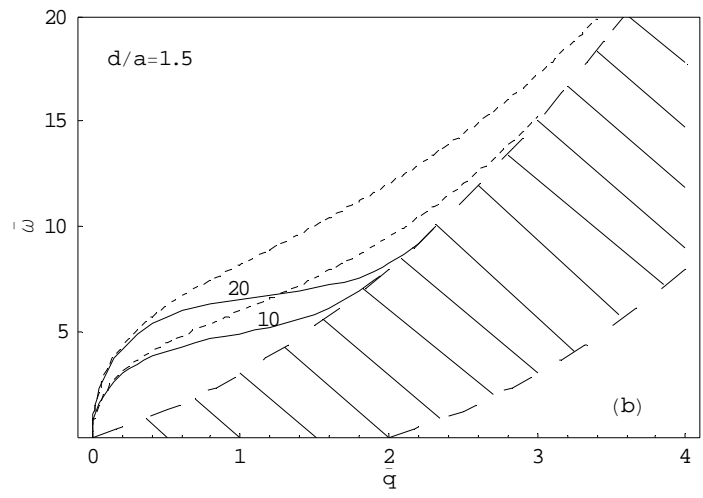

FIGURE 13

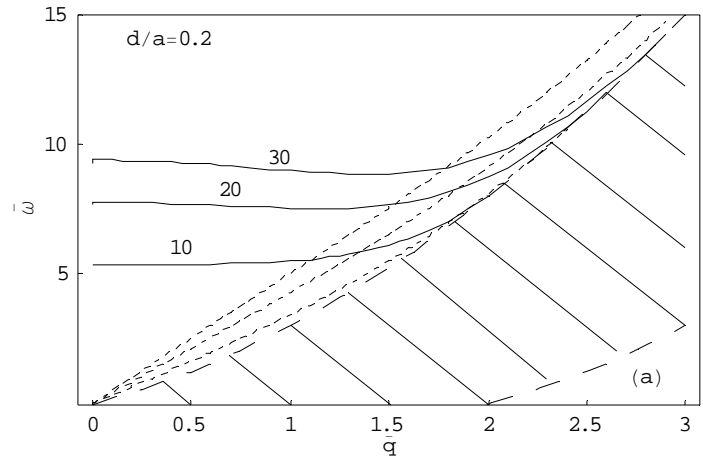
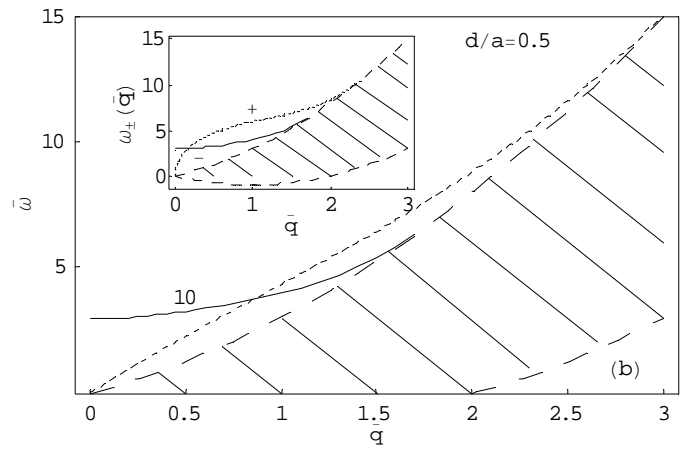

FIGURE 14

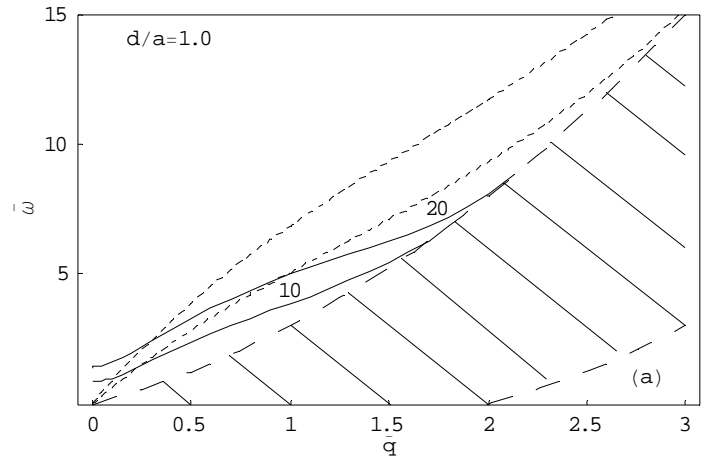

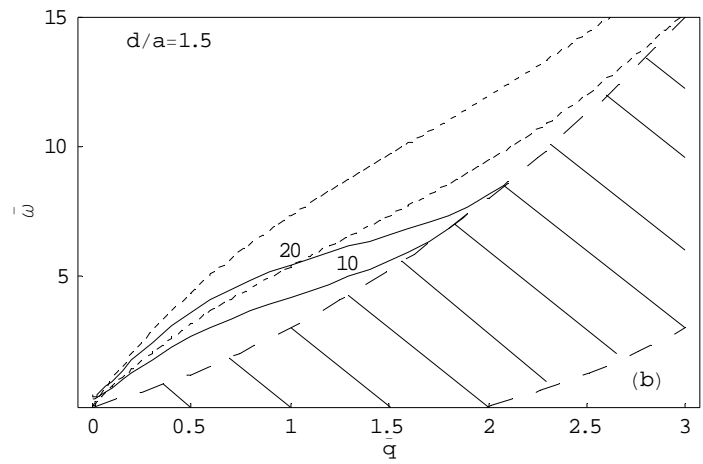

FIGURE 15